\documentclass{article}% for LaTeX 2e
 \usepackage{graphics} % or 
\usepackage{epsfig }
\begin{document}

\title{Single hadron transverse spin
asymmetries from COMPASS}

\author{Franco Bradamante~\footnote{
Presented at 17th International Spin Physics Symposium (SPIN06),
        2-7 October 2006, Kyoto, Japan,
Proceedings to be published from AIP, American Institute of Physics.}\\
 University of Trieste and INFN Trieste \vspace*{0.5cm}\\
 {\em on behalf of the COMPASS Collaboration}}
\date{ }
\maketitle

\begin{abstract}
Transverse spin physics is an important part of the scientific
programme of the COMPASS experiment at CERN.
The analysis of the data taken with the target polarized 
orthogonally to the 160 GeV/c muon beam momentum has allowed to measure
for the first time the Collins and Sivers asymmetries
of the deuteron.
Both for the positive and the negative hadrons produced
in semi-inclusive DIS the measured asymmetries are small
and, within errors, compatible with zero.
New results for $\pi^{\pm}$ ans K$^{\pm}$ are presented here.
\end{abstract}
{\small
\begin{description}
\item[{\rm {\em PACS:}}] 13.60.-r, 13.88.+e, 14.20.Dh, 14.65.-q
\item[{\rm {\em Key words:}}] transverse spin effects, parton distribution functions,
                semi-inclusive deep inelastic scattering
\end{description}
  }

\vspace*{1cm}
The COMPASS experiment has measured for the first time single 
hadron transverse spin asymmetries in DIS of high energy muons on deuterons,
scattering the 160 GeV/c muon beam at the CERN SPS on a
transversely polarised $^6$LiD taget.
In
such processes the measurable asymmetry $A_{Coll}$ (``Collins asymmetry'')
is due to the combined effect of the transversity distribution function (DF)
 $\Delta_T q(x)$ and another chirally-odd function, $\Delta^0_T D_q^h$, which
describes the spin dependent part of the
hadronization of a transversely polarized quark $q$ in
a hadron $h$.
At leading order 
$A_{Coll}$ can be written as 
\begin{eqnarray}
A_{Coll} = \frac {\sum_q e_q^2 \cdot \Delta_T q \cdot \Delta_T^0 D_q^h}
{\sum_q e_q^2 \cdot q \cdot D_q^h}
\label{eq:collass}
\end{eqnarray}
where $e_q$ is the quark charge.
The quantities $\Delta_T^0 D_q^h$ can be obtained by
investigating the fragmentation of a polarised quark
$q$ into a hadron $h$, f.i. in $e^+e^- \rightarrow$ hadrons.

A different mechanism has also been suggested in the past as a 
possible cause of
a spin asymmetry  in the cross-section of SIDIS 
between leptons and transversely polarised nucleons.
Allowing for an intrinsic $k_T$ dependence of the quark distribution
in a nucleon, a left-right asymmetry could be induced in such a distribution
by the transverse nucleon polarisation, thus causing an asymmetry 
$A_{Siv}$ (the ``Sivers
asymmetry'') in the quark fragmentation hadron with respect to the nucleon 
polarisation.
\begin{eqnarray}
A_{Siv} & = & \frac {\sum_q e_q^2 \cdot \Delta_0^T q \cdot D^h_q}
{\sum_q e_q^2 \cdot q \cdot D_q^h} 
\label{eq:sivass}
\end{eqnarray}

The asymmetries $A_{Coll}$ and $A_{Siv}$ can be extracted separately 
from semi-inclusive hadron production in DIS scattering on transversely 
polarised nucleons by measuring a $\sin \Phi_C$ or $\sin \Phi_S$
modulation in the azimuthal distributions of the hadrons,
where $\Phi_C=\phi_h+\phi_S-\pi$ and $\Phi_S=\phi_h-\phi_S$
\begin{figure}
\begin{center}
  \includegraphics[width=.65\textwidth]{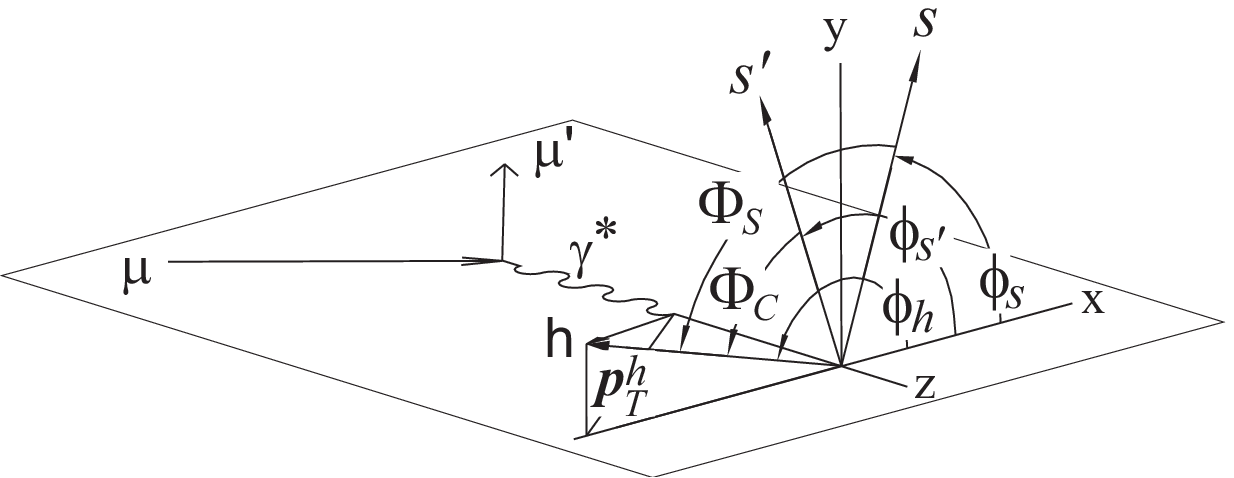}
  \caption{Definition of the Collins and Sivers angles.}
\label{fig:angle}
\end{center}
\end{figure}
(see Fig.~\ref{fig:angle} for the definition of the reference system).

The COMPASS experiment is described in these Proceedings~\cite{Magnon} and in more
detail in Ref.~\cite{NPB}. 
About 20\% of the total beam-time in 2002, 2003 
and 2004 was devoted to the run 
with the transversely polarised deuteron target. 
The kinematic cuts 
$Q^2>1$ (GeV/c)$^2$, $W > 5$ GeV/c$^2$ and $0.1< y <0.9$ were 
applied to the data, where
$Q^2$ is the photon virtuality, $W$ the mass of the
hadronic state, and $y$ the fractional energy of the virtual photon.
The energy fraction of the hadron, $z$, was required to be larger
than 0.2 for the ``all hadron'' sample, and 0.25 for the 
``leading hadron'' sample.
After all the analysis cuts the total number of DIS events was 
about 10$\cdot 10^6$.

\begin{figure}
\begin{center}
  \includegraphics[width=.78\textwidth]{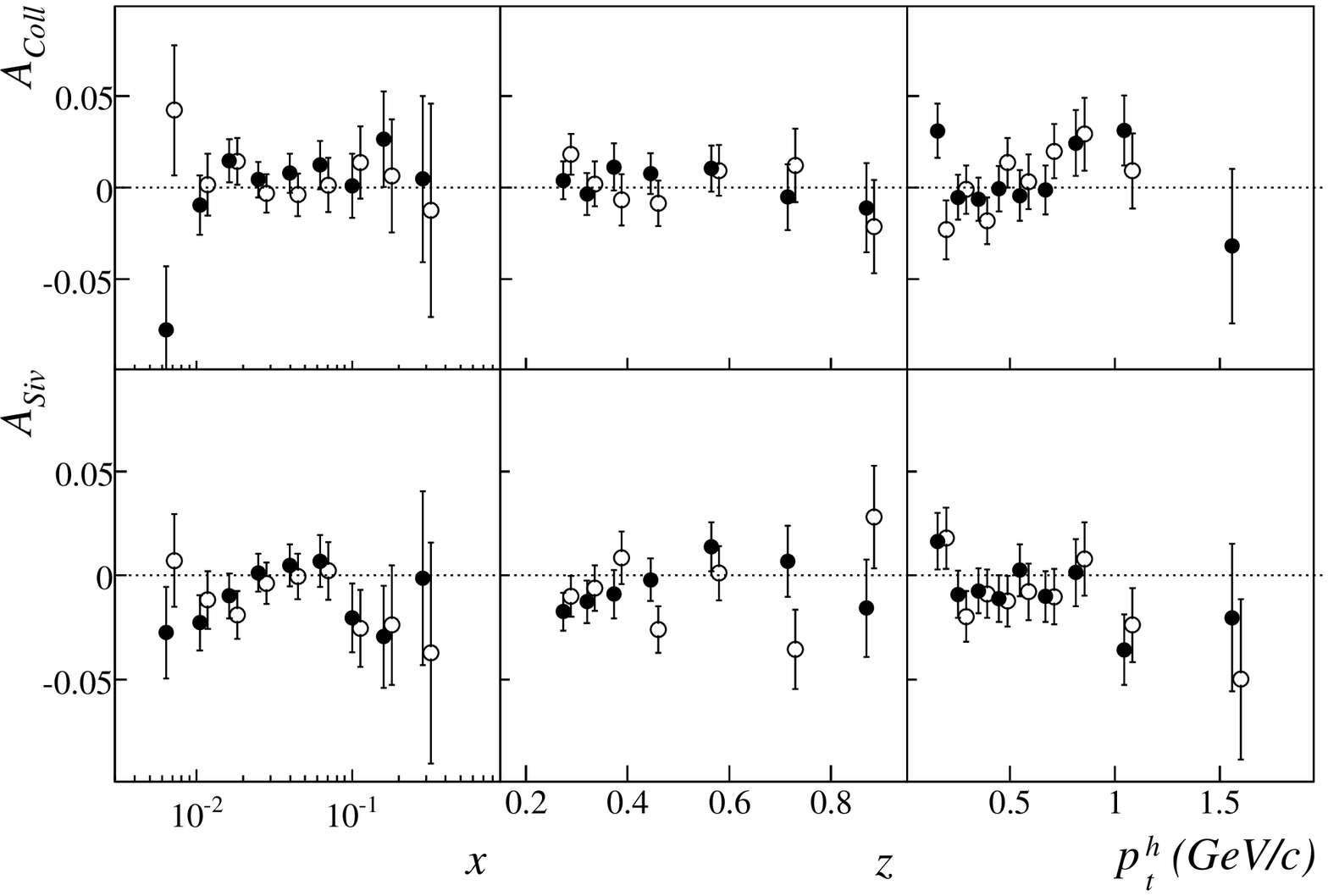}
  \caption{Preliminary results for Collins asymmetry (top) and Sivers asymmetry 
(bottom) against $x$, $z$ and 
$p_T^h$ for ``leading'' $\pi^+$ (full circles) and ``leading'' $\pi^-$
(open circles) from 2003 and 2004 data.}
\label{fig:cspi.eps}
\end{center}
\end{figure}

\begin{figure}
\begin{center}
  \includegraphics[width=.78\textwidth]{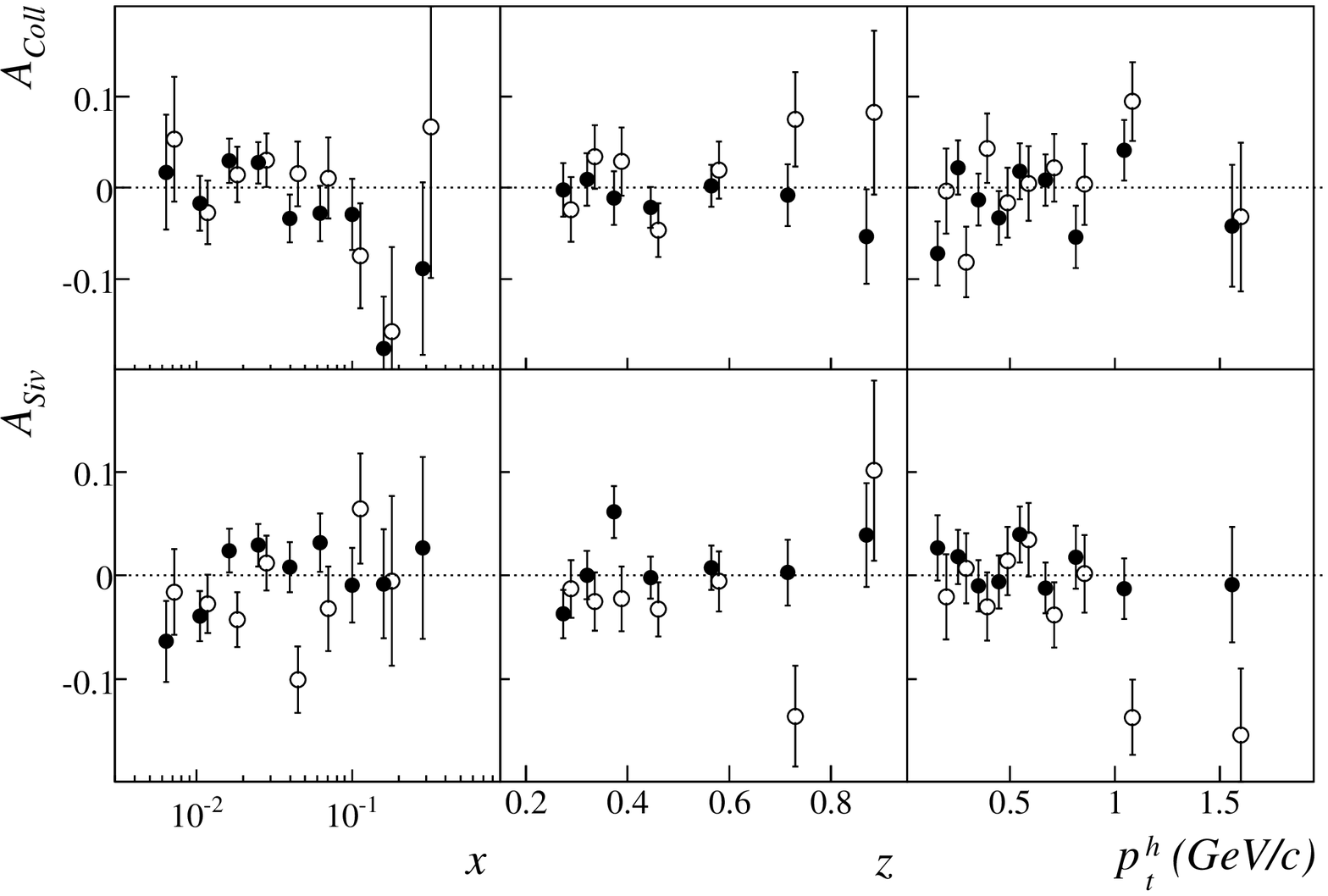}
  \caption{Preliminary results for Collins asymmetry (top) and Sivers asymmetry 
(bottom) against $x$, $z$ and 
$p_T^h$ for ``leading'' K$^+$ (full circles) and ``leading'' K$^-$
(open circles) from 2003 and 2004 data.}
\label{fig:csk.eps}
\end{center}
\end{figure}

The resulting Collins and Sivers asymmetries 
for ``all'' and ``leading'' hadrons from the 2002, 2003, and 2004 data  
are final.
They   have already been
accepted for publication~\cite{NPB},
and are shown in~\cite{Magnon}.
The preliminary results for the $\pi^{\pm}$ and  K$^{\pm}$ asymmetries 
are plotted against the kinematic variables 
$x$, $z$ and $p_T$ 
in Fig.~\ref{fig:cspi.eps} for ``leading'' $\pi^{\pm}$
and in Fig.~\ref{fig:csk.eps} for ``leading'' K$^{\pm}$.
They refer to the 2003 and 2004 
data, and correspond to $6.2 \cdot 10^6$
and $1.1 \cdot 10^6$  events for $\pi$'s and K's
respectively.
Full points correspond to the positively charged hadrons and 
open points correspond to the negatively charged hadrons.
Only statistical errors are shown:
systematic errors have been shown to be considerably
smaller  than statistical ones.

As apparent from Fig.~\ref{fig:cspi.eps} and \ref{fig:csk.eps}, 
all the measured asymmetries are small, 
if any, and compatible with zero. 
This trend already characterised the published data of the 2002 
run~\cite{ourPRL}, 
and is confirmed by the new data with considerably improved precision. 

An analysis of the results on the deuteron can be done only in conjunction 
with corresponding proton data, which up to now have been measured only 
by the HERMES Collaboration~\cite{Hermest}. 
Within the  statistical accuracy of the  HERMES 2002-2004 data on 
protons and our 2002 data on deuterons,
a few global analysis aiming at the extraction
of the Sivers functions and of the transversity distributions
have already been performed, and the observed 
phenomena can be described in a unified 
scheme~\cite{thwork}.

Independently from the model calculations, I'd like to stress that we
are witnessing the discovery of new phenomena.
The measured non zero Collins asymmetry on the proton
 has provided convincing evidence that both the transversity distribution 
$\Delta _T u(x)$ and
 the Collins mechanism $\Delta_T ^0 D_u^h (z)$ are not zero
(the present HERMES data has allowed
 to extract only the leading contribution to the proton transverse asymmetry, 
that is the u quark contribution).
Independent evidence that the Collins mechanism is a real 
measurable effect has come from the recent analysis of the
BELLE Collaboration~\cite{Belle}. 
Furthermore, the HERMES data on a proton target have provided
convincing evidence that the Sivers  mechanism is also at work.

In the following,  naive expectations
are given for the pion asymmetries, 
which are valid also for the non-identified hadrons, since about 80\% of them
are pions. Formulas (\ref{eq:collass}) and (\ref{eq:sivass}) simplify considerably
by neglecting 
the sea contribution and considering only the valence $x$-region,
i.e. the region where the HERMES and COMPASS data overlap,
and the HERMES data show non-zero values.
Assuming: 
$D_u^{\pi^+}  =  D_d^{\pi^-} =  D_1 , \; \; 
D_d^{\pi^+}   =  D_u^{\pi^-} = D_2 , \;\; 
\Delta_{T}^0 D_u^{\pi^+}  = \Delta_{T}^0 D_d^{\pi^-} =  \Delta_{T}^0 D_1   , \;\; 
\Delta_{T}^0 D_d^{\pi^+}  =    \Delta_{T}^0 D_u^{\pi^-} = \Delta_{T}^0 D_2$,
and using Eq.~(\ref{eq:collass}), one gets for a deuteron target
\begin{equation}
A_{Coll}^{d, \pi^+}  \simeq  
\frac{\Delta_{T} u_v  + \Delta_{T} d_v}{u_v + d_v }
\frac{4 \Delta_{T}^0 D_1 + \Delta_{T}^0 D_2}
     {4 D_1 + D_2} \,  , \; \;
A_{Coll}^{d, \pi^-}  \simeq 
\frac{\Delta_{T} u_v  + \Delta_{T} d_v}{u_v + d_v }
\frac{\Delta_{T}^0 D_1 + 4 \Delta_{T}^0 D_2}{D_1 + 4 D_2} \, .
\label{eq:colpid1b}
\end{equation}

The fragmentation term is known to be different
from zero (and HERMES data suggest that
$ \Delta_{T}^0 D_1 \simeq -  \Delta_{T}^0 D_2$),
therefore the smallness of both the $\pi^+$ and $\pi^-$
Collins asymmetries we have measured on the deuteron is a 
first indication that 
$ \Delta_{T} u_v  \simeq -  \Delta_{T} d_v$.
One has to stress that in so far the model calculations
which have analysed our partial results from the 2002 data
were not able to constrain the transversity of the
d-quark, but this should be possible now, with our new
precise data.

Also for the Sivers asymmetry it is useful to consider the expressions one obtains 
for the pions.
Again, the simplified analysis neglects the sea contribution,
and is restricted to the valence region.
For a deuteron target the Sivers asymmetries can be written as
\begin{equation}
A_{Siv}^{d, \pi^+}  \simeq   
\frac{\Delta_0^{T} u_v  + \Delta_0^{T} d_v}{u_v + d_v}   \,  , \; \;
A_{Siv}^{d, \pi^-}  \simeq 
\frac{\Delta_0^{T} u_v  + \Delta_0^{T} d_v}{u_v + d_v}
\label{eq:sivpid1b}
\end{equation}
which implies $A_{Siv}^{d, \pi^+}  \simeq   A_{Siv}^{d, \pi^-}$.
The approximatively zero Sivers asymmetries for positive and negative
hadrons observed in COMPASS require
$
\Delta_0^{T} d_v \simeq -  \Delta_0^{T} u_v \, ,
$
a relation which is also obtained in some models,
and which anyhow has a simple physical interpretation if the Sivers
distortion of the PDF of the nucleon is associated with the
orbital angular momentum of the u and d quarks,
whose sum almost vanishes in the deuteron.
As a matter of fact,
if we accept that in the deuteron the contributions of the u and the d
quark should cancel,
the smallness of the Sivers asymmetry for positive and 
negative hadrons on the deuteron can be interpreted as evidence for
the absence of gluon orbital angular momentum in the 
nucleon, as suggested in Ref.~\cite{Brodsky:2006ha}.

%\begin{theacknowledgments}
%  ....
%\end{theacknowledgments}



\bibliographystyle{aipproc}   % if natbib is available

\end{document}